# Nanoscale chemical engineering of $Ti_3C_2T_x$ MXene multilayers through *in situ* electron beam irradiation in a TEM

*Ayoub Benmoumen[1,2§], Marie-Laure David[1]*, Noé Condé[1], Eric Gautron[2], Sophie Morisset[3], Stéphane Célérier[3], Anupma Thakur[4], Babak Anasori[4,5], Vincent Mauchamp[1]*, Philippe Moreau[2]**

[1] Université de Poitiers, CNRS, ISAE-ENSMA, PPRIME, Poitiers, France

[2] Nantes Université, CNRS, Institut des Matériaux de Nantes Jean Rouxel, IMN, F-44000 Nantes, France

[3] Université de Poitiers, CNRS, Institut de Chimie des Milieux et Matériaux de Poitiers (IC2MP), 86073 Poitiers, France

[4] School of Materials Engineering, Purdue University, West Lafayette, IN 47907, United States

[5] School of Mechanical Engineering, Purdue University, West Lafayette, IN 47907, United States

ABSTRACT

Defect engineering in low-dimensional materials is a powerful approach to tailor their structure and chemistry at the atomic scale, with high relevance for tuning physical properties. Here, we apply this strategy to MXenes, one of the largest families of two-dimensional (2D) materials, and demonstrate the use of a highly focused scanning transmission electron microscope (STEM) beam for the structural and chemical nanoengineering of titanium carbide ($Ti_3C_2T_x$) multilayers. By adjusting the beam energy, electron dose, and scanning parameters, $Ti_3C_2T_x$ multilayers were irradiated at 80 and 300 keV and characterized *in situ* using annular dark-field imaging and electron energy-loss spectroscopy (EELS). The irradiation induces gradual structural and chemical modifications, largely driven by the selective sputtering of atomic species and MXene layers. We identify irradiation conditions that selectively remove Cl or F surface terminations while largely preserving the $Ti_3C_2$ framework, whereas other conditions promote preferential carbon removal, in agreement with theoretical predictions. Most notably, electron irradiation drives the migration and accumulation of initially dilute nitrogen impurities within the irradiated regions, leading to the formation of nitrogen-rich titanium carbonitride ($Ti_3C_yN_zT_x$) nanodomains embedded in the $Ti_3C_2T_x$ multilayers. These laterally patterned carbide/carbonitride architectures constitute in-plane MXene heterostructures with nanometer-scale spatial control. These results demonstrate that electron irradiation can be used to locally tailor MXene composition and create nanoscale architectures beyond those accessible through conventional chemical synthesis.

## 1. Introduction

Low-dimensional (low-d) materials and their assembled nanostructures exhibit extraordinary electrical, optical, and mechanical properties that arise from their reduced dimensionality and strong quantum confinement effects. These characteristics make them ideal candidates for key sectors of modern society, such as energy storage and conversion, nanoelectronics, optoelectronics, catalysis, or quantum technologies[1–4]. Beyond demonstrated potential in advanced device fabrication and functional materials design, the possibilities offered by low-d systems can be dramatically enhanced by tailoring their structure and composition with ultimate atomic precision, opening avenues in properties tuning. This can be achieved at different levels, including controlled defect formation (vacancies and heteroatom doping)[5–7], optimized layer stacking[8], or interlayer spacing[9], among others. In this context, electron beam sculpting has emerged as a powerful approach for such nano to atomic-scale design. By exploiting the highly focused and tunable electron probe in a (scanning) transmission electron microscope (S)-TEM, it is possible to locally remove, rearrange, or deposit matter with sub-nanometer precision while simultaneously imaging the process at play in real time, a key advantage for the optimization of this nanostructuration strategy.

The electron beam of a (S)-TEM interacts with matter through elastic and inelastic processes, generating a variety of defects within the layers, due to atomic sputtering or displacements and to electronic excitations[10]. The type and distribution of the as-created structural defects depend on several factors, including the atomic species involved, their position within the lattice and the irradiation parameters, namely, electron energy, dose, dose rate and sample temperature. Such controlled alterations can modulate the electronic structure, chemical reactivity, and mechanical behavior of nanomaterials, enabling customized functionalities.

Recent advances have enabled the sculpting of various geometries and the improvement of the understanding of electron-matter interaction in a range of low-d materials, from graphene[11] and hexagonal boron nitride[12] to transition metal dichalcogenides[13] and nanowires[14]. These studies have revealed new insights into atomic diffusion, defect dynamics, and the stability of low-d systems under electron irradiation. For instance, the synthesis of metallic nanoparticles by electron beam-induced fragmentation or nucleation/precipitation has been widely reported, as well as topological modifications leading to two-dimensional (2D) nanostructures[15]. Other examples include control of plasmonic structures[16], building of structures atom by atom[17], sculping of graphene[18], nanomaterials nucleation and growth[19], drilling of solid state nanopores[20], or control of helium density in helium nanobubbles[21].

2D materials are one of the building pillars in the development of low-d systems, and 2D transition metal carbides, nitrides, and carbonitrides, known as MXenes, are among the latest and fastest growing family[22]. Since their first synthesis in 2011[23], they have shown outstanding potential for various fields of applications[22] thanks to the combination of high electrical conductivity[24], a large range of optical properties[25], remarkable chemical diversity[22], tunable hydrophilicity, and easy and cost-effective processing[26]. MXenes have been investigated for many different applications, including optoelectronics[27,28], energy storage and conversion[29,30], biomedicine[31,32], catalysis[33], sensors[34,35], or EMI shielding[36,37]. These 2D layers are mostly obtained from the selective chemical etching of the A layers from a very large family of three dimensional (3D) nanolaminated precursors known as MAX phases, and whose general formula is $M_{n+1}AX_n$, where M is a transition metal, A usually belongs to groups III-A and IV-A of the periodic table, X is C and/or N and $n$ is usually ranging from 1 to 4[38]. This directly determines MXene chemical formula, which can be written as $M_{n+1}X_nT_x$ ($x \approx 2$), where T refers to surface functional groups inherited from the exfoliation process (typically, -F, =O, -OH or -Cl)[39]. Each structural component of MXene multilayers—*i.e.*, the $M_{n+1}X_n$ core, the surface terminations, and intercalants—plays a decisive role in determining their physical properties[25,40–44]. Because of these multiple levels of tunability, MXene 2D layers have become a versatile platform where solid solution effects and defects[7,45], surface functionalization control[41], or interlayer engineering[40] can be largely used to enhance or modulate properties over a broad spectrum.

In the present work, we further expand structural and chemical engineering strategies for MXenes by exploring controlled electron-beam irradiation of $Ti_3C_2T_x$ multilayers in the STEM as a route to nanoscale composition and structure control. Our study builds upon recent theoretical and experimental studies of electron–MXene interactions. *Ab initio* molecular dynamics (AIMD) simulations have provided valuable insights into the sputtering behavior of $Ti_3C_2T_x$ monolayers, revealing the sputtering thresholds of various surface groups[46]. Experimentally, several studies have demonstrated that MXenes can be manipulated at the nanoscale by electron irradiation. Zhang et al. reported the redistribution of MX core atoms and surface functional groups in $Ti_3C_2T_x$ under electron-beam exposure,[47] while Sang et al. demonstrated controlled structural modification of $Ti_3C_2T_x$ monolayers leading to the homoepitaxial growth of single-layer TiC flakes at elevated temperature.[48] More recently, Boebinger et al. demonstrated the controlled formation of atomically defined nanopores,[20] and nanoscale rippling has also been reported in $Ti_3C_2T_x$ MXenes with organic surface terminations under low-temperature electron irradiation.[49]

Here, we show that the structure and chemistry of $Ti_3C_2T_x$ MXenes can be selectively modified by tuning the electron-beam conditions. Using aberration-corrected STEM, we systematically vary the acceleration voltage and electron dose and follow the resulting transformations by annular dark-field (ADF) imaging and spatially resolved electron energy-loss spectroscopy (EELS). Cl surface terminations can be preferentially removed while largely preserving the $Ti_3C_2$ framework, whereas higher doses promote carbon depletion and layer removal. The response strongly depends on the electron energy, in agreement with calculated sputtering thresholds. Most notably, irradiation-induced carbon depletion is accompanied by the migration and local accumulation of initially dilute nitrogen impurities, enabling the formation of nitrogen-rich $Ti_3C_yN_zT_x$ nanodomains with nanoscale lateral control. These results extend electron-beam sculpting of MXenes from structural modification to local chemical design and demonstrate a route toward chemically patterned carbide/carbonitride nanoarchitectures.

## 2. Methods

### *2.1. Sample preparation*

$Ti_3C_2T_x$ MXene flakes were synthesized from $Ti_3AlC_2$ MAX phase precursors using an LiF/HCl solution in MILD (minimally intensive layer delamination) conditions and following the protocol detailed in M. Benchakar *et al.*[50]. The powder was dispersed in ultra-pure water first by a simple hand shaking, followed by sonification for 5 minutes. The aqueous solution was then drop casted on a lacey carbon copper grid.

$Ti_3C_yN_zT_x$ ($y \approx 1$, $z \approx 1$) MXene was synthesized using wet-chemical selective etching of $Ti_3AlC_yN_z$ MAX using a mixed acid approach. For synthesis, 1 g of the synthesized $Ti_3AlC_yN_z$ MAX was mixed with an etchant solution comprising 12 M HCl, deionized water, and 28.4 M HF in volumetric ratios of 6:3:1. The reaction mixtures were stirred at 400 RPM for 24 h at 35 °C. The etched multilayered $Ti_3C_yN_zT_x$ MXene was washed with deionized water via repeated centrifugation at 3234 RCF (500 mL of deionized water) until the supernatant reached pH ∼6. The aqueous solution of multilayered $Ti_3C_yN_{yz}T_x$ MXene was then drop-cast on a lacey carbon copper grid.

### *2.2. STEM and EELS Acquisition conditions*

The STEM measurements were performed using a probe-corrected Themis Z G3 from Thermo Fisher Scientific, operated at 80 and 300 kV accelerating voltage, at liquid nitrogen temperature using a Fischione cryogenic sample holder (Model 2550). To limit ice formation on the sample during the low temperature experiments, the TEM grid was placed on the sample holder tip between two aluminium plaques. These plaques were drilled to obtain a Ø200 µm hole allowing the electron beam to pass through, while preventing completely ice accumulation (see SI 5).

Electron Energy-Loss Spectroscopy spectrum imaging (EELS-SI) was performed in microprobe mode with a 0.69 mrad convergence semi-angle, a 3.6 mrad collection semi-angle and a probe size of 3.2 nm full width at tenth maximum at 300 kV. At 80 kV, the values are respectively 1 mrad, 3 mrad and 1.4 nm. The rather large probe size minimizes the electron dose, safely out of the beam damage range; hence, local chemical analysis can be performed with a dose calculated to be ≈ $1.2 \times 10^5$ $e^-/Å^2$ (probe current around 150 pA at 300 kV), much lower than those used for irradiation engineering.[51] The current and the exposition time were chosen in such a way to prevent the sample from electron beam degradation, following recently published simulations showing that the overall surface composition of $Ti_3C_2T_x$ MXene flakes should not be altered in such conditions[46]. The spectra were recorded on a GIF Quantum ERS 966 spectrometer (Gatan), equipped with a K2 Summit direct electron detection camera (Gatan) that was used for core loss measurements. The thickness t of the areas studied was measured using a US1000 CCD camera (Gatan). The t/λ ratio (λ, inelastic mean free path) was computed from the low loss measurements using the log-ratio method[52,53]. The thickness was then calculated assuming Malis'[52] formula for the inelastic mean free path shown to give reasonable results for sufficiently thick $Ti_3C_2T_x$ multilayers[54]. In our experimental conditions, the inelastic mean free path was estimated at around 168 nm for 300 keV and around 113 nm for 80 keV.

*2.3. In situ electron beam irradiation protocol*

The electron irradiation experiments were performed using a protocol we refer to as "high dose spectrum-image irradiation". It consists in using a typical STEM-SI acquisition, while significantly increasing the electron current (up to 1.5 nA) and exposure time per pixel (up to 35 s/px). The spectrum images used were typically 4×4 or 5×5, with 5 nm pixel size. This can result in the accumulation of a very high electron dose at each pixel of the SI (up to $1.22 \times 10^9$ $e^-/Å^2$ in this study). By continuously tuning the electron current and exposure time, we are able to choose accurately the electron dose received by the MXene stack, allowing us to follow gradually the evolution of the sample as a function of the electron dose. This method could generate an extremely high intensity, which would permanently damage the camera. To avoid this, we shift the energy range detected by the camera to very high values (> 3000 eV), where the signal is extremely weak, thus avoiding the over-saturation of the detector. Once irradiations are performed, the microscope settings can be switched back to typical acquisition conditions and core loss spectra can be obtained. EELS quantifications were performed on raw spectra using the DigitalMicrograph (DM) software (Gatan), by adjusting the fitting parameters for the pristine sample, then applying the same fit on all the remaining spectra to avoid user bias. These same parameters were also used to obtain the chemical maps, using the "elemental mapping" option in DM, giving us the signal (electron count) maps shown in figures 4. Assuming a typical 10% error on the elemental quantification of each element, the error on the X/Ti ratio (X ≡ Cl, C, N) is estimated to be around 20%.

*2.4. Diffraction patterns simulations*

Theoretical electron diffraction patterns were obtained by all beams dynamical calculation using the Blochwave method including experimental parameters (convergence semi-angle, accelerating voltage) using the JEMS software.[55]

*2.5. X-ray photoelectron spectroscopy*

The XPS measurements shown in SI were performed on a Kratos Axis Ultra using a monochromatic X-ray Al Kα source (1486.6 eV) at 20 mA and 15 kV. The analyzed surface was about 700 μm by 300 μm and the instrument base pressure was $5 \times 10^{-9}$ Torr. All samples were maintained on carbon conductive adhesive double-tape. A pass energy (PE) of 80 eV was used to acquire wide-range survey spectra with a resolution of 0.9 eV for metallic silver. A PE of 40 eV was used to acquire narrow spectra of the different core levels with a resolution of 0.55 eV measured on the Fermi edge of the valence band for metallic silver. Charge compensation has been used for every measurement. XPS data were analyzed using the CasaXPS software (version 2.3.27) using Gaussian/Lorentzian line shapes and Shirley background correction. All spectra were calibrated with hydrocarbon C 1s photoemission set to 285.0 eV binding energy. The C 1s energy corresponding to the carbon peak in the $Ti_3C_2T_x$ MXene is consequently found at 281.9 eV, in agreement with literature.

**3. Results and discussion**

$Ti_3C_2T_x$ MXene powders were synthesized using the LiF + HCl etching method starting from $Ti_3AlC_2$ MAX phases[50,56], and resulting in well delaminated flakes after dispersion in water. MXene flakes were drop-casted on a lacey TEM copper grid (**Figure 1a**), and the *in situ* electron beam irradiation experiments were performed using a S/TEM operated at 80 or 300 keV. To achieve better control over the irradiation process and to mitigate electron beam-induced contamination, the $Ti_3C_2T_x$ samples were irradiated at low temperature using a liquid-nitrogen-cooled sample holder (T ≈ 96 K). Preliminary analysis of the $Ti_3C_2T_x$ MXene stacks was performed before irradiation, first using selected area electron diffraction (SAED) to verify their crystallographic quality. The ring-like pattern shown in **Figure 1b** is a signature of a $Ti_3C_2T_x$ MXene flake with the layers' basal planes randomly oriented, with no sign of oxidation or degradation being observed. EELS measurements were also performed, giving the typical energy-loss spectrum shown in **Figure 1c**. This spectrum contains the core edges of all the elements present in the sample: the carbon K- and titanium $L_{2,3}$-edges characteristic of the MX core, the chlorine $L_{2,3}$-, oxygen K- and fluorine K- edges corresponding to the surface functionalization groups, and finally the nitrogen K-edge. This element is present in the samples as a low concentration impurity at a level of ≈ 1 – 2 at% according to X-ray photoelectron spectroscopy (XPS) and an elemental analysis performed with a CHNS/O Analyzer (see SI 1). The source of this impurity was traced back to the MAX phase precursor.

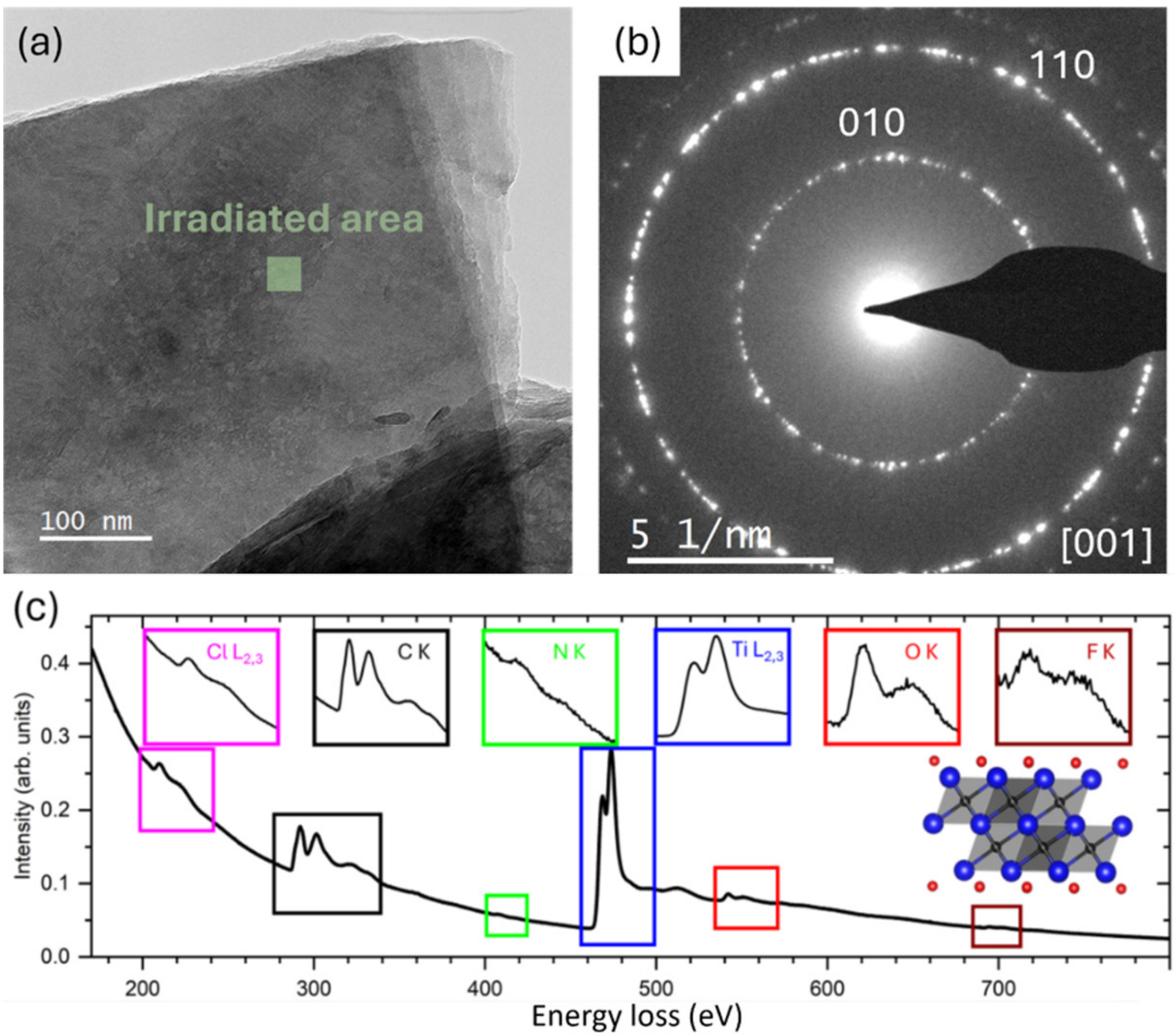

**Figure 1:** (a) TEM micrograph of a $Ti_3C_2T_x$ multilayer, with the corresponding selected area electron diffraction pattern (b). (c) Typical EEL core loss spectrum for pristine $Ti_3C_2T_x$. Insets: the Cl $L_{2,3}$-, C K-, N K-, Ti $L_{2,3}$-, O K-, and F K-edges. The presented spectra are not background-

subtracted. A schematic structure of a $Ti_3C_2T_x$ MXene sheet is also given: titanium atoms are represented in blue, carbon atoms in black, and surface groups in red.

**Table 1.** Summary of the electron doses corresponding to the irradiation experiments at 300 and 80 keV

| | 300 keV | 80 keV |
|---|---|---|
| | (x $10^7$ $e^-/Å^2$) | |
| $d_1$ | 6.12 | 20.3 |
| $d_2$ | 7.41 | 40.6 |
| $d_3$ | 11.1 | 81.2 |
| $d_4$ | 22.2 | 122 |
| $d_5$ | 41.5 | - |

Well-defined nanometer-scale regions of the samples (≈ 25 x 25 $nm^2$) were irradiated under various conditions in STEM mode, using high electron doses of up to 1.22 x $10^9$ $e^-/Å^2$. Details of the irradiation protocol are provided in the Methods Section. The irradiations were performed with electron beam energies of 300 and 80 keV. The electron doses used for both acceleration voltages, labelled $d_1$ to $d_5$, are summarized in **Table 1**. It should be noted that, compared with 300 keV, substantially higher electron doses are required at 80 keV to induce measurable structural or chemical modifications.

The outcomes of these experiments are shown in **Figures 2** and **3**, which illustrate the effects of irradiation at increasing electron doses. The morphological and chemical evolutions of the samples were monitored *in situ* in the TEM. A first stack, 55 nm thick (*i.e.,* about 50 $Ti_3C_2T_x$ layers), was irradiated with an electron beam at 300 keV, using the five electron doses reported in **Table 1** (*i.e.*, from 6.12 to 41.5 × $10^7$ $e^-/Å^2$). The corresponding ADF micrograph is given in **Figure 2a**. A second stack (11 nm thick) was subjected to irradiation at 80 keV, with electron doses in the range 20.3 × $10^7$ to 122 × $10^7$ $e^-/Å^2$ (see **Table 1**), the corresponding ADF micrograph being given in **Figure 2c**. The formation of a clear defect pattern is observed in both cases for the highest electron doses, as evidenced by the darker contrast in the ADF micrographs in **Figure 2a** and **2c**, especially for areas $d_4$ and $d_5$ at 300 keV and $d_3$ and $d_4$ at 80 keV. Only slight changes in contrast are observed for the other areas. This trend is further highlighted by the line profiles shown in **Figures 2b** and **2d**, which reveal a gradual increase in the depth of the sputtering-induced dip as a function of the electron dose for both experiments. For the 300 keV irradiations, we also notice a sharp decrease in the ADF intensity between $d_3$ and $d_4$. This pronounced intensity difference is only observed in this case, even though the electron dose is also almost doubled between $d_1$ and $d_3$, as well as between $d_4$ and $d_5$. This could imply that a threshold was reached between 11.1 × $10^7$ $e^-/Å^2$ and 22.2 × $10^7$ $e^-/Å^2$, where a significant number of defects were created, weakening the MXene layers and resulting in a faster displacement of atoms in the sample. A similar behavior was observed by Boebinger *et al.*[20] while irradiating $Ti_3C_2$ monolayers. They observed the formation of an intermediate metastable $Ti_2C$ phase, which withstands electron irradiation until a threshold electron dose is accumulated, causing the collapse of the structure and full sputtering of the sample.

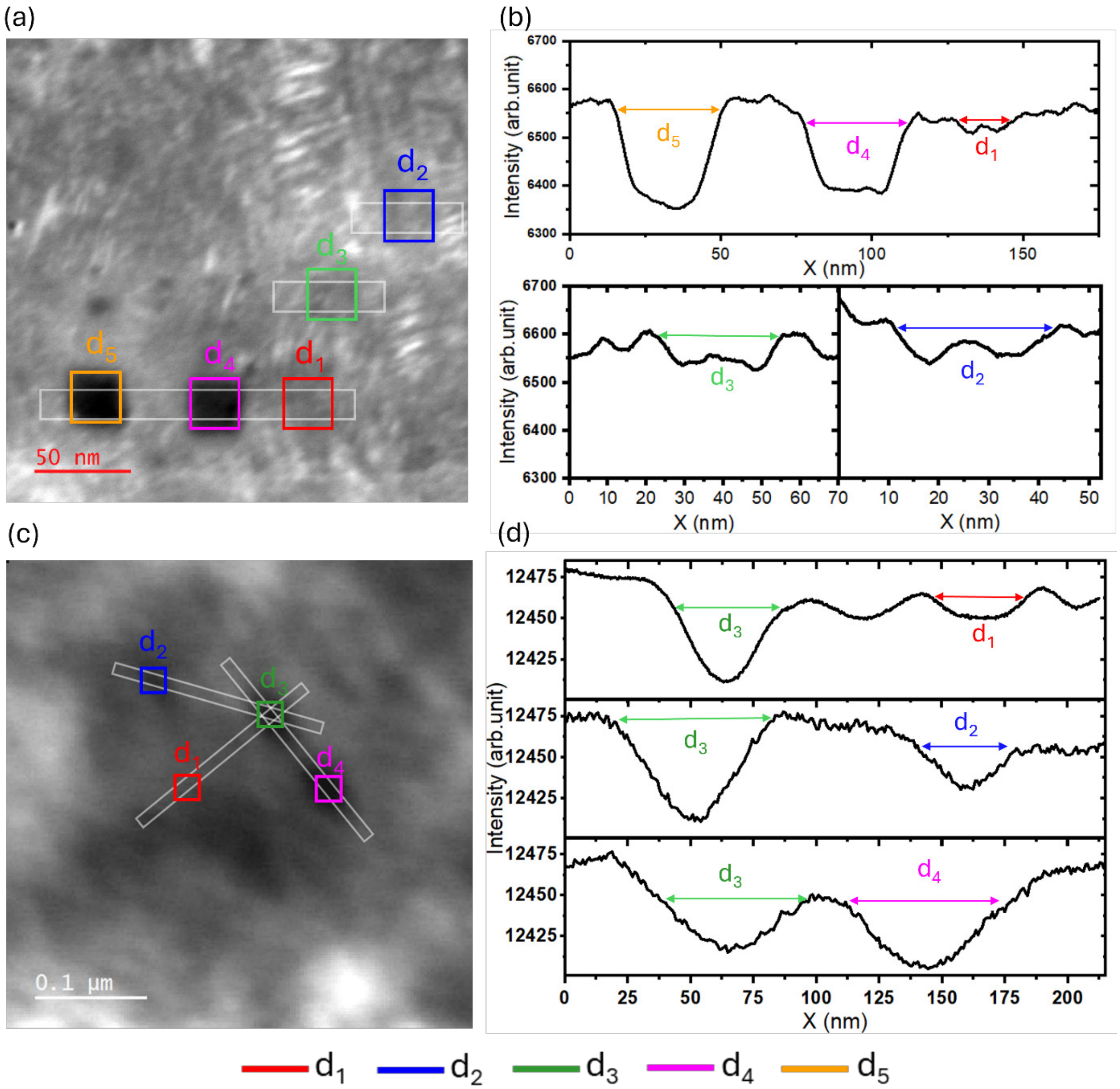

**Figure 2:** (a,c) ADF micrographs showing the areas of the $Ti_3C_2T_x$ MXene flakes irradiated at 300 keV (a) and 80 keV (c). The colored squares represent the region of interest (ROI) taken for each dose (see legend at the bottom of the figure). (b,d) Line profiles over the irradiated areas in the micrographs, integrated in the white rectangles in (a,c), and showing the depth and size of the resulting dips for 300 keV (b) and 80 keV electron irradiations (d).

Focusing now on the shape of the irradiated areas, a thickness increase was systematically observed on the edges of the damaged zones, as evidenced by an increase in the ADF intensity.

This implies the redeposition of displaced atoms in the proximity of the ROI edges, increasing the mass-thickness signal. A similar behavior was reported in the literature[20,48], where a pileup of Ti atoms was observed around nanopores created in monolayer $Ti_3C_2T_x$ using electron beam irradiation. This phenomenon was also confirmed by simulations[20,48]. One can also observe a bump in the middle of areas 1-3 for the 300 keV irradiation, with a deeper sputtering in the edges. This is attributed to the scanning method of the beam, a raster scan, which causes a higher accumulation of the electron dose at the edges while the system recalibrates to travel back to the beginning of the next scan line. This results in a higher electron dose and higher irradiation at the edges.

Although a similar irradiation protocol was used for both energies, the formation of the nanopores occurs differently: sharp edges are observed at 300 keV, following the spectrum image shape, while a gradual pore formation towards the center of the spectrum image is observed at 80 keV (**Figure 2d**). This change can be attributed to the different sputtering rates of the atoms at the two energies, especially for Ti, for which the sputtering threshold is only reached for 270 keV electrons according to DFT calculations.[46] Overall, our findings validate the feasibility of creating nanopores in multilayer $Ti_3C_2T_x$ MXene via electron beam irradiation, under carefully controlled experimental conditions. However, the primary objective of this study is to investigate the structural and chemical modifications induced in the remaining layers when complete sputtering is not achieved.

Using low electron dose acquisition conditions ($\approx 7 \times 10^5$ $e^-/Å^2$) to avoid further atom displacement, STEM-EELS analyses were thus performed on the irradiated areas to study the chemical composition of these remaining layers, starting with the functional groups. We focus on the Cl $L_{2,3}$-edge as it gives the strongest EELS signal among T-groups and can be quite straightforwardly extracted from the background compared to the more challenging O K- and F

K-edges. The Cl $L_{2,3}$-edges of the pristine and irradiated areas are given in **Figures 3a** and **3b** for 300 keV and 80 keV electron irradiations, respectively. For 300 keV, we observe only a minor decrease in the EELS intensity for the first three electron doses, and then a sudden drop for the fourth and fifth doses, in correlation with the atomic displacement threshold discussed in **Figure 2b**. As for the 80 keV irradiation, a more gradual decrease of the EELS intensity is observed as a function of the electron dose. This trend is also reflected in the Cl/Ti ratios (**Figure 3c**), obtained from elemental quantifications of the spectra given in **Figures 3a** and **3b** for 300 keV and 80 keV, respectively. The data were normalized to the pristine sample value for better clarity.

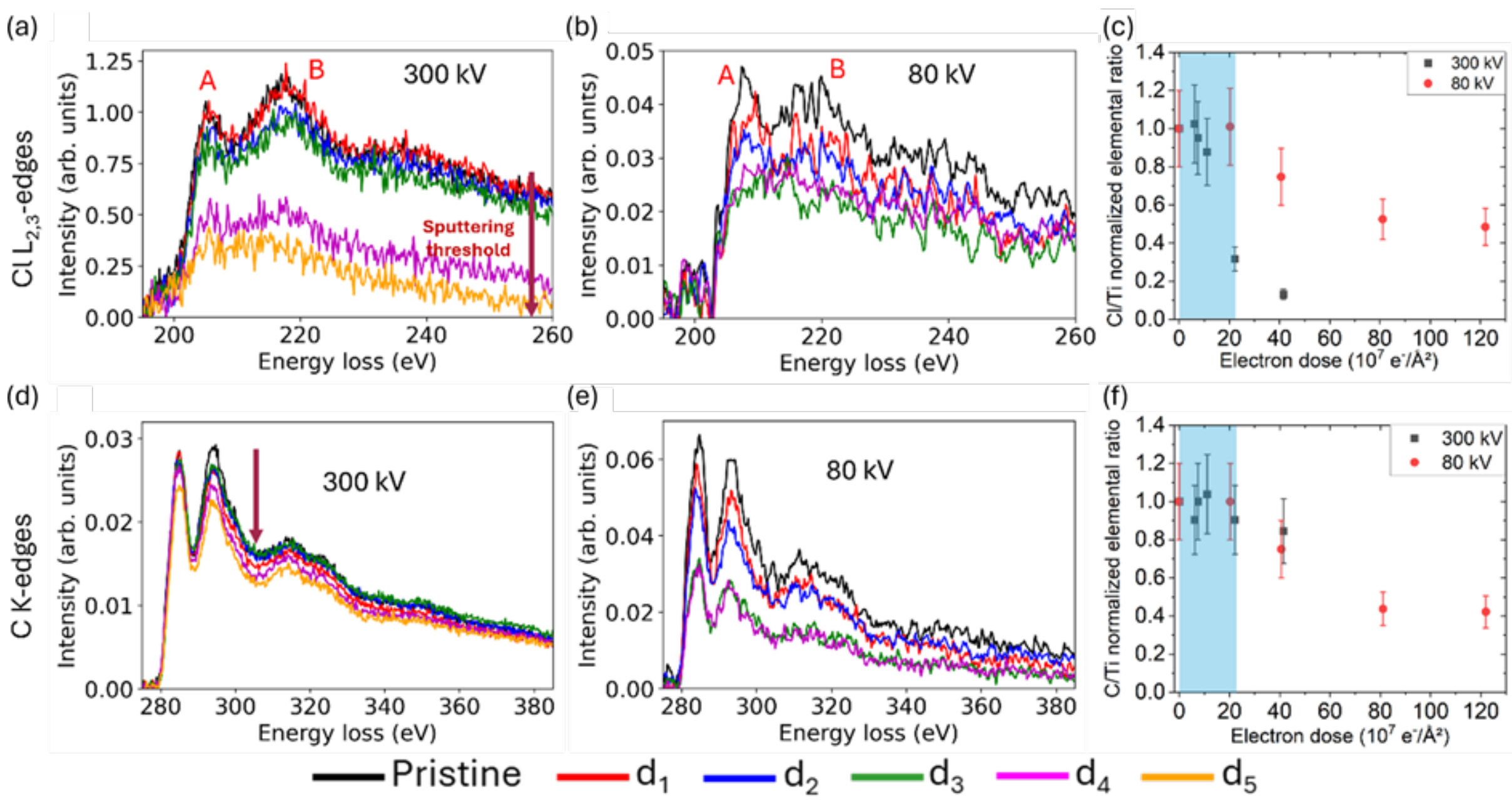

**Figure 3:** (a,b) Cl $L_{2,3}$-edges measured on pristine and irradiated areas of the $Ti_3C_2T_x$ MXene sample after the 300 keV (a) and 80 keV (b) electron irradiations. (c) Normalized elemental ratio of Cl/Ti as a function of the electron dose. (d,e) C K-edges measured on pristine and irradiated areas of the sample at 300 keV (d) and 80 keV (e). (f) Normalized elemental ratio of C/Ti as a function of the electron dose. The EEL spectra were normalized to the Ti $L_{2,3}$-edge intensity to highlight intensity changes relative to the Ti content in the corresponding areas. Assuming a typical 10% error on the Cl, C and Ti contents, the error on the Cl/Ti and C/Ti ratio is estimated to be around 20%.

A sharp drop in the Cl/Ti ratio occurs at 300 keV between $d_3$ and $d_4$, reaching approximately 13% of its initial value at $d_5$. In contrast, a more progressive decrease is observed at 80 keV, where the ratio reaches about 52% of its initial value for $d_4$, even though this electron dose is three times higher than the maximum one obtained at 300 keV. In both cases, we observe a preferential removal of Cl compared to Ti, indicating that this functional group is more easily removed, with the removal rate increasing strongly with the electron energy. One can also observe that the fine structures at the Cl $L_{2,3}$-edge slightly change upon departure of Cl (**Figure 3a),** with an inversion of the intensity ratio between peaks A and B after $d_4$. Given the sensitivity of the fine structure at the T-groups edges to their local environment[57,58], this reveals a noticeable change in the local environment around Cl atoms after $d_3$, consistent with the threshold already discussed before.

The analysis for the F K-edge is more delicate since the edge is poorly defined in our experimental conditions. This is particularly true at 80 keV, where this edge is barely visible even in the pristine sample – see SI 2. One can, however, notice that the F K-edge intensity relative to the Ti $L_{2,3}$-edge clearly decreases after $d_4$ and $d_5$ for the 300 keV irradiation, as evidenced by the spectra given in SI 2, indicating a strong removal of these functional groups, consistent with our observations for Cl-groups. Regarding the O K-edge, no clear changes are observed between the pristine and irradiated areas (SI 2a and 2b), regardless of the electron beam energy. The fine structures as well as the overall edge intensity, relative to the Ti $L_{2,3}$-edge, are stable, which indicates little change in the O content. The sputtering cross section calculations for O (SI 3) suggests that O is prone to sputtering for both 80 and 300 keV electron energies, and a decrease in the O/Ti might therefore be expected under both irradiation conditions. The observed stability of this ratio in these experiments could be attributed either to the recombination of displaced O with active defects in the layers or to external O sources such as interlayer water.

As for the MX core, the C K-edges of the irradiated areas are presented in **Figures 3d** and **3e** for 300 keV and 80 keV irradiation, respectively. First, no signs of amorphization or strong changes in the fine structure were noticed, indicating that the as-deposited $Ti_3C_2T_x$ MXene structure remains fairly stable even in the highly irradiated areas. As for the carbon concentration, a slight decrease of the signal is observed at 300 keV, suggesting a mild removal relative to titanium (spectra are normalized to the Ti signal) as the electron dose increases. The signal shows a more significant decrease under 80 keV electrons, meaning a stronger relative removal of carbon in this case. This is further quantified thanks to the normalized C/Ti ratios shown in Figure 3f. Under 300 keV electrons, only a small decrease of about 15% is observed for $d_5$, whereas under 80 keV electrons, a significantly larger decrease of about 58% is calculated at dose $d_4$. The preferential sputtering of carbon compared to titanium at 80 keV and high electron doses is in accordance with the theoretical calculations, since titanium is predicted to be stable under the electron beam whereas the sputtering cross section for carbon reaches almost its highest value (see SI 3). At 300 keV the situation is different since titanium begins to be sputtered while the sputtering cross section for carbon decreases, leading to a lower relative sputtering of C compared to Ti, a result fully consistent with our experimental observations.

The different sputtering rates of Cl and C relative to Ti at 300 keV open a window in the irradiation conditions (highlighted in blue in **Figures 3c** and **3f**) that enables the selective removal of Cl-terminations of the MXene without substantially altering the $Ti_3C_2$ core chemical composition. This is interesting since Cl terminal groups were reported to be highly stable under annealing, and thus hard to remove from MXene surfaces[59], making electron irradiation a promising approach to this end. In this context, using electron beam irradiation on MXenes fully (or mainly) terminated with Cl, as obtained from molten salt approaches[60,61], appears as an interesting approach to reach bare (unterminated) MXene surfaces with potentially interesting properties arising from their highly reactive surface (*e.g.*, $CO_2$ capture[62] or work function control through termination[63,64]).

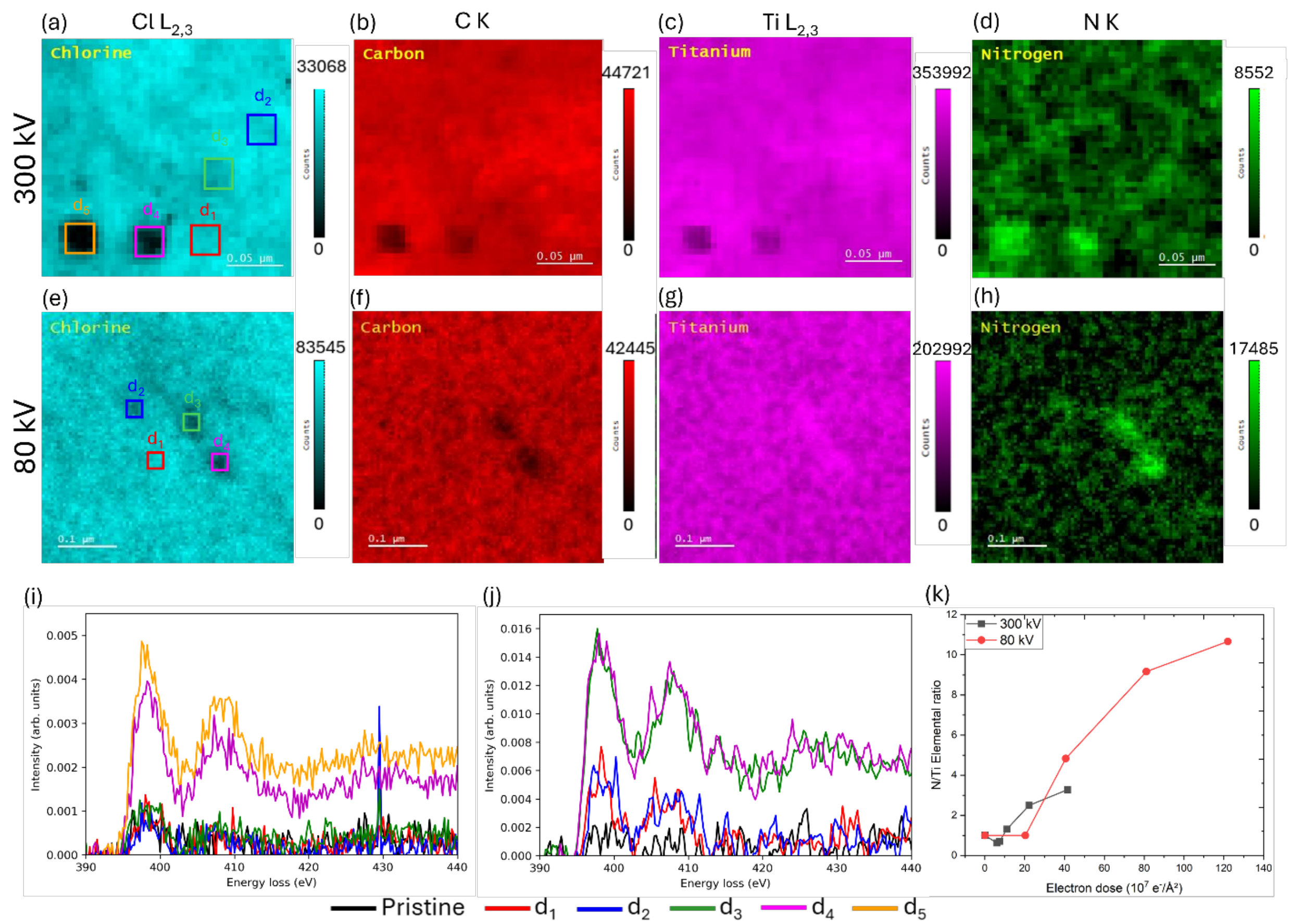

**Figure 4:** (a-h) Signal maps obtained from the raw data of the EELS spectrum images of the studied areas for $Ti_3C_2T_x$ MXene at 300 keV (a-d) and 80 keV (e-h) (corresponding to the irradiated areas ($d_1$ to $d_5$) shown in Figure 2a and 2c). N K-edges measured on pristine and irradiated areas of the sample at 300 keV (i) and 80 keV (j). (k) Normalized elemental ratio of N/Ti as a function of the electron dose. The EEL spectra were normalized to Ti $L_{2,3}$-edge intensity to highlight intensity changes relative to the Ti content in the corresponding area. Assuming a typical 10% error on the N and Ti contents, the error on the N/Ti ratio is estimated to be around 20%.

Beyond these results, electron irradiation reveals an additional, unexpected, and remarkably inspiring result, as evidenced by the elemental signal maps of the irradiated areas given in **Figure 4**. These maps give the net recorded intensities (*i.e.*, the electron count) of the irradiated areas at 300 keV (**Figures 4a-d**) and 80 keV (**Figures 4e-h**). They first clearly confirm the intensity decrease for the Cl, C and Ti signal in the irradiated areas at 300 keV, indicating a removal of all of these elements, and a preferential removal of carbon and chlorine atoms as compared to titanium atoms in the areas irradiated at 80 keV (**Figure 4g**). More interesting is the opposite effect observed for the nitrogen signal, which increases dramatically in the irradiated areas (**Figures 4d** and **4h**). As already mentioned, nitrogen was only present as an impurity in the pristine samples, spread around homogeneously in the layers. This significant accumulation of N in the irradiated areas was consistently observed throughout all our experiments and is very reproducible. At 300 keV, barely any change in the intensity of the N K-edge is observed for $d_1$ to $d_3$ **(Figure 4i)**. However, a sudden increase is observed for $d_4$ and $d_5$, in line with the threshold effect mentioned previously. As for the 80 keV electron irradiation, the nitrogen signal increases with the electron dose, showing a slight increase for $d_1$ and $d_2$, and a more significant increase for $d_3$ and $d_4$ (**Figure 4j**). The N-to-Ti intensity ratio also appears to increase with the electron dose under both irradiation conditions (see SI 4). These evolutions are further highlighted in the elemental quantification results shown in **Figure 4k**, where the N/Ti ratio is plotted as a function of the electron dose. After a rather constant profile for the lowest doses, a sharp increase is observed for both irradiation energies, with an approximately three-fold maximum increase of the nitrogen concentration at 300 keV and an impressive tenfold maximum increase at 80 keV. These chemical modifications were obtained on square regions of about 25×25 $nm^2$ but can easily be reproduced on larger areas or areas with different shapes, thanks to the precise control of the electron beam in modern electron microscopes.

The comparison between the evolution of the C/Ti and N/Ti ratios (**Figures 5a** and **5b**) reveals an interesting anti-correlation, where the carbon sputtering is mirrored by the nitrogen accumulation in the irradiated areas. As MXenes can be carbonitrides, it is reasonable to assume that a substantial part of the accumulated nitrogen atoms can replace the sputtered carbon atoms in the octahedral sites. This hypothesis is further reinforced by the elemental quantification results, where the (C+N)/Ti ratio of the irradiated areas appears to progress towards the expected value of approximately 0.66 for an ideal $Ti_3X_2$ system ($X = C_yN_z$) for the 300 keV irradiation (**Figure 5a**). For the 80 keV irradiation (**Figure 5b**), the combined ratio starts at the expected value of around 0.66 and further increases as a function of the electron dose up to around 0.9 ($y+z \geq 2$), leading to an oversaturation of nitrogen, where it eventually exceeds the amount of carbon in these areas. The resulting nano-domains can therefore be described as nitogen-rich MXenes. A plausible explanation is that nitrogen atoms occupy both some of the carbon sites within the lattice and the surface sites left vacant by sputtered termination groups. Notably, DFT calculations have predicted that nitrogen can occupy surface termination sites on the MXene surface.[65] In addition, XPS analysis given in SI 1 shows two distinct features in the N 1s peak of the $Ti_3C_yN_zT_x$ MXene, confirming that N occupies at least two sites in the MXene structure.

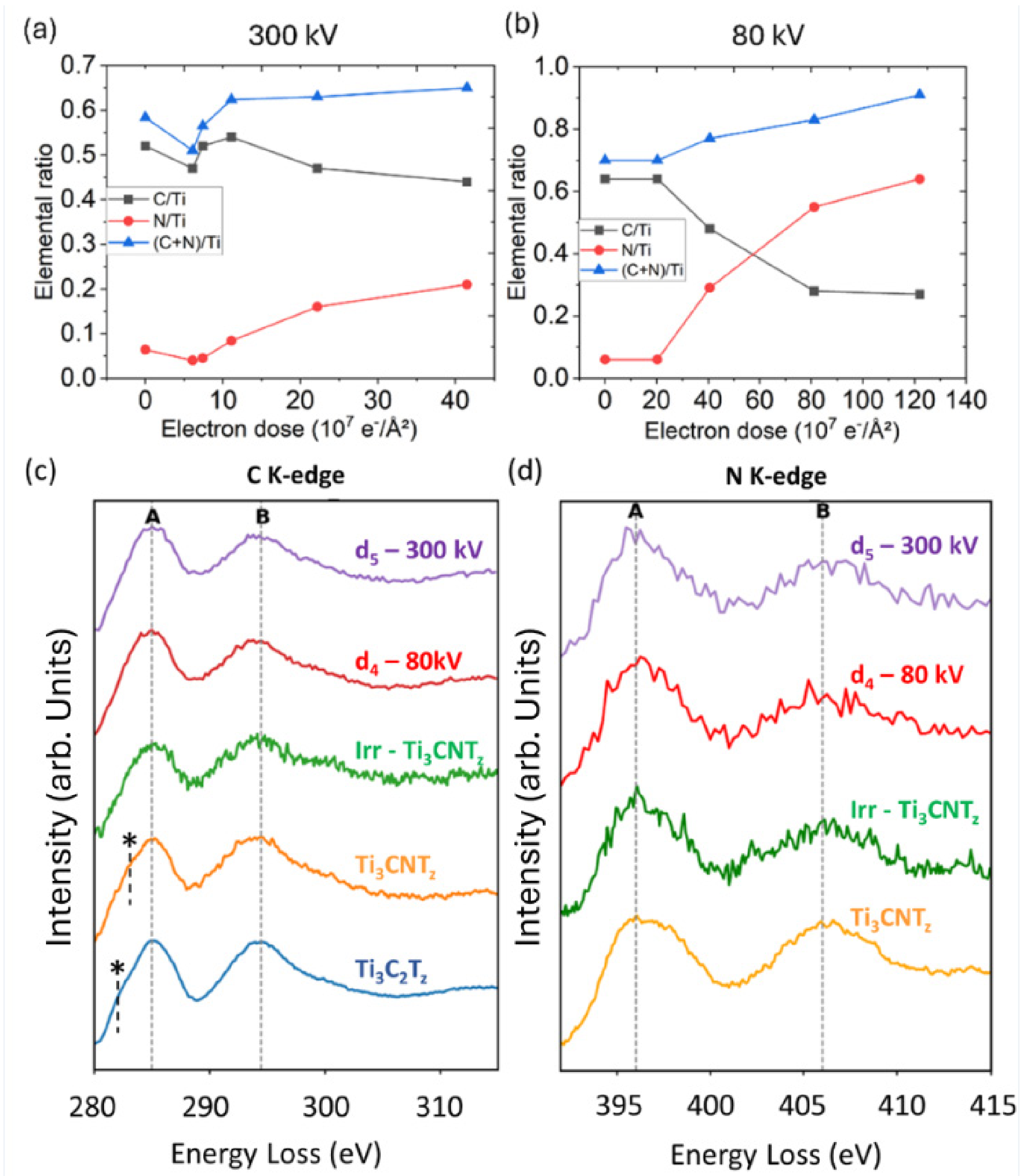

**Figure 5:** (a,b) C/Ti (black curve), N/Ti (red curve), and (C+N)/Ti (blue curve) elemental ratios in $Ti_3C_yN_zT_x$ MXene samples as a function of the electron dose at 300 keV (a) and 80 keV (b). Assuming a typical 10% error on the C, N and Ti contents, the error on the C/Ti and N/Ti ratios is estimated to be around 20%. (c) Comparison between the C K-edge recorded on pristine $Ti_3C_2T_x$, pristine $Ti_3C_yN_zT_x$, irradiated $Ti_3C_yN_zT_x$, $d_4$-80 keV, and $d_5$-300 keV (from bottom to top). (d) Comparison between the N K-edge recorded on pristine $Ti_3C_yN_zT_x$, irradiated $Ti_3C_yN_zT_x$, $d_4$-80 keV, and $d_5$-300 keV (from bottom to top).

To better understand the impact of electron irradiation on the local atomic arrangement in $Ti_3C_2T_x$, we synthesized a $Ti_3C_yN_zT_x$ (y ≈ 1, z ≈ 1) carbonitride (see the Methods section for the description of the synthesis) and conducted electron irradiation on this carbonitride MXene multilayers. The corresponding irradiation experiments were performed in a JEOL 2200 FS equipped with an in-column omega filter, and thus in different conditions from those used previously: parallel illumination with a 200 keV electron beam and at room temperature. The electron dose was not precisely determined. Our purpose here was to compare the C and N K-edges in well-crystallized $Ti_3C_yN_zT_x$ flakes and corresponding irradiated layers with the data collected on the irradiated carbide layers. A similar sputtering effect as for $Ti_3C_2T_x$ under STEM irradiation conditions was observed, and the thickness of the irradiated area, as derived from EELS measurements, decreased from 58 to 33 nm during the exposure, confirming similar behaviour to that obtained for $Ti_3C_2T_x$.

The C K-edges recorded in the carbide and carbonitride MXenes before and after electron irradiation are compared in **Figure 5c**. In the pristine materials, the edges look quite similar with two main peaks labelled A and B. As already documented in the literature[66], the peak A contains different substructures, identified by stars in the figure, which are slightly different in $Ti_3C_yN_zT_x$ and $Ti_3C_2T_x$ (different energy positions, different intensities), reflecting a slightly different local environment for carbon atoms when nitrogen is incorporated in the structure. After electron irradiation, the introduced disorder induces a global broadening of both peaks A and B, leading to a very similar shape between the irradiated carbonitride and the spectra collected from the nitrogen-rich areas in $Ti_3C_2T_x$ at $d_4$-80 keV and $d_5$-300 keV electron irradiations. One can, however, notice that the relative intensity of peak A vs peak B in the irradiated carbonitride is a bit smaller than in the irradiated carbide: this could well result from slightly different surface chemistries of the layers according to the analysis of T. Bilyk *et al.*[66]. Similar comparisons were also performed at the N K-edge, the corresponding results being given in **Figure 5d**. Notably, the N K-edge fine structures of the nitrogen-containing nano-domains in the carbide MXene exhibit features closely matching those of the irradiated carbonitride, suggesting again a comparable local atomic environment in the two irradiated MXenes. Altogether, these findings suggest that the local environment of carbon and nitrogen atoms in the irradiated $Ti_3C_2T_x$ layers is probably very comparable to that in a disordered $Ti_3C_yN_zT_x$ MXenes, featuring variable and potentially high nitrogen concentrations with a controlled nanometer-scale size distribution.

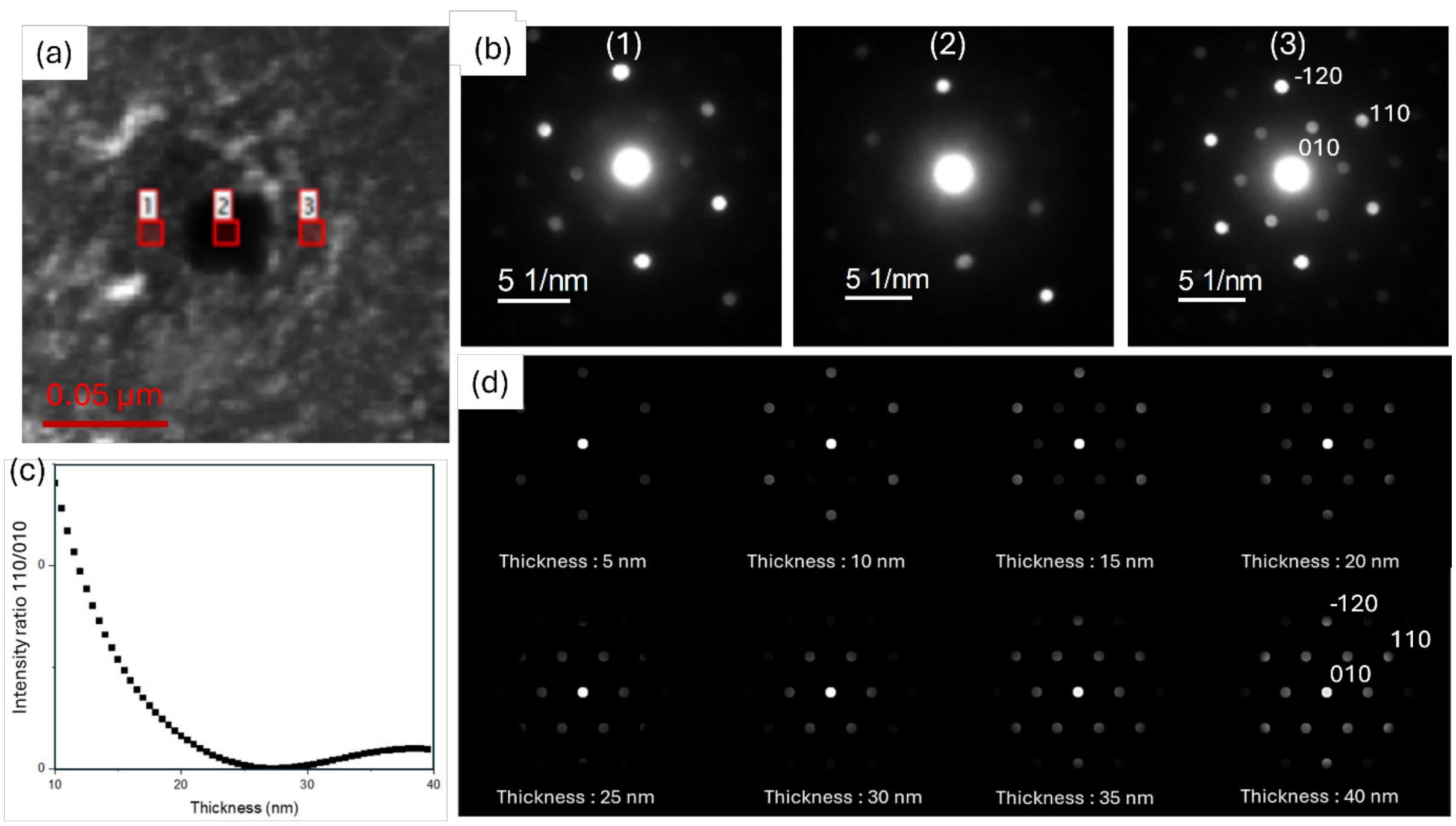

**Figure 6:** (a) ADF micrograph of a $Ti_3C_2T_x$ multilayer. The red rectangles indicate the areas where the diffraction patterns in (b) were taken. (b) Electron diffraction patterns extracted from 4D-STEM data, the left (1) and right (3) ones being outside of the irradiated area, while the middle one (2) is inside the irradiated area. These experiments were performed on a 33 nm thick flake irradiated at 300 keV with an electron dose of $21 \times 10^7$ $e^-/Å^2$. (c) Intensity ratio of the 110 and 010 diffraction spots as a function of the MXene stack thickness, obtained from JEMS simulations.[55] (d) JEMS simulations of electron diffraction patterns of $Ti_3C_2F_2$ multilayers at thicknesses ranging

from 5 to 40 nm, and highlighting the evolution of the intensity of diffraction spots as a function of the thickness.

The crystallographic structure of the nitrogen-rich areas in $Ti_3C_2T_x$ is quite close to that of the pristine MXene, as confirmed by 4D STEM experiments (**Figure 6)**. These experiments were performed on a 33 nm thick flake irradiated at 300 keV with an electron dose of $21 \times 10^7$ $e^-/Å^2$. Diffraction patterns corresponding to non-irradiated parts of the layer are compared to a diffraction pattern collected in the irradiated zone (**Figure 6b**). All spots corresponding to [001] orientation are clearly visible in the pristine sample. After the 300 keV electron irradiation, while the 110-type reflections are still visible, the 100-type spots are no longer revealed. Very importantly, no other diffraction spots, which would be characteristic of a second phase formation during irradiation, are observed, and one should notice that the diffraction pattern of the pristine $Ti_3C_yN_zT_x$ sample is very similar to that of $Ti_3C_2T_x$. The disappearance of the 100 spots can be ascribed to dynamical and/or structural effects.

JEMS simulations[55] indicate that the ratio of the 100- to 110-type spot intensities decreases significantly with decreasing sample thickness, from 27 nm down to a few layers (**Figures 6c** and **6d**). Consequently, sputtering-induced thinning may partially explain why the 100-type spots are barely visible after irradiation. In addition, the presence of defects, including titanium atoms located within the surface termination layers[47], the formation of nanoripples[49], or localized disorientations, could also contribute to modifying the diffraction pattern. Further investigations are required to test these hypotheses. Note that the disappearance of the 100-type reflections has been reported previously after *in situ* electron irradiation of the $Ti_3AlC_2$ parent MAX phase[67]. Overall, these data are fully in line with the EELS results discussed previously: the crystallographic structure of the irradiated layers is expected to be quite similar to that of the pristine layer.

The reason why N is not removed by the electron beam but rather clustered into the irradiated area is not clear at the moment. A similar behavior was however observed in other materials, providing some clues about the potential mechanisms at play in our experiments. In the work of V. Garcia *et al.*[68], the authors observed, using Auger electron spectroscopy measurements, a migration of nitrogen atoms under electron beam irradiation in silicon dioxide thin films previously nitrided in a plasma environment. Their explanation is based on the migration of the ionized nitrogen atoms towards the electron beam because of the electric field it generates while irradiating the sample. This phenomenon would be similar to that commonly referred to as "contamination", when accumulation of amorphous C is observed under the electron beam. The second hypothesis assumes that this phenomenon is caused by the so-called "*radiation induced segregation*" (RIS) process. It is typically invoked in the case of ion implantation experiments or strong irradiation settings (nuclear reactors, for example), where the impurities in the sample segregate towards the areas with high vacancy concentration[69–71]. N. Moncoffre and J. Jagielski [72] used this process to explain the migration of implanted nitrogen in iron towards the surface at high temperatures, proving that nitrogen (as an impurity) tends to diffuse towards areas with high defect concentration at high temperatures. This could be a reasonable explanation for the nitrogen accumulation observed in our irradiated samples, since we observed an accumulation of nitrogen in strongly irradiated areas. Even if thermal activation is highly unlikely in our experimental conditions, atom diffusion on large scales seems indeed to occur in our case. Finally, one could notice that the higher displacement resistance of N compared to C could be attributed to the reported stronger Ti – N bond compared with Ti – C[73].

## 4. Conclusion

In this study, controlled *in situ* electron irradiation in the STEM is used to chemically engineer $Ti_3C_2T_x$ MXene multilayers at the nanoscale. By systematically varying the electron energy and dose and monitoring the resulting evolution with ADF imaging and spatially resolved EELS, we identify distinct regimes of structural and chemical transformation. At 300 keV, a dose-dependent sputtering threshold is observed, whereas the evolution is more gradual at 80 keV. Surface terminations, particularly Cl and F, are preferentially removed compared with Ti, and Cl depletion can be achieved under conditions that largely preserve the $Ti_3C_2$ framework, providing a route toward locally unterminated $Ti_3C_2$ surfaces. At higher doses, carbon is preferentially removed relative to titanium, particularly at 80 keV, in agreement with calculated sputtering thresholds. These results demonstrate that selective control of the M, X, or T components of the MXene can be achieved by adjusting the irradiation conditions.

A particularly striking result is the formation of nitrogen-rich nanodomains within $Ti_3C_2T_x$ multilayers. Starting from materials containing only ≈1–2 at% nitrogen impurities, irradiation induces nitrogen migration and accumulation within the exposed regions, leading to $Ti_3C_yN_zT_x$ nanodomains with controlled lateral dimensions and spatial distribution. TEM characterization shows that these regions retain a structure close to that of the pristine layers, supporting the formation of N-rich MXene-like domains. Although the mechanisms governing nitrogen redistribution remain to be fully elucidated, this result demonstrates that irradiation-induced defects can be exploited not only to remove atomic species but also to redistribute minor elements and generate locally modified compositions.

More broadly, these results establish that a highly focused electron probe can be used to locally tailor their composition and create new nanoscale architectures. The carbide/carbonitride nanostructures demonstrated here illustrate the potential of beam-driven chemical engineering for creating laterally patterned MXene architectures that would be impossible to obtain by conventional synthesis. Although the relevant irradiation thresholds will depend on MXene composition and thickness, the underlying strategy should be applicable to other MXenes and potentially to other low-dimensional materials. This opens a route toward spatially controlled chemical modification and nanoscale heterostructures with tailored composition and properties.

ASSOCIATED CONTENT

**Supporting Information**.

SI 1 - XPS and CHNS/O Analyzer measurements of nitrogen impurity concentrations. SI 2 - O K-, F K- and Ti $L_{2,3}$-edges of pristine and irradiated areas at 300 kV and 80 kV, normalized to the Ti $L_{2,3}$-edge intensity. SI 3 - Sputtering cross-section calculations. N K- and Ti $L_{2,3}$-edges of pristine and irradiated areas at 300 kV and 80 kV, normalized to the Ti $L_{2,3}$-edge intensity. SI 4- N K- and Ti $L_{2,3}$-edges of pristine and irradiated areas at 300 kV and 80 kV, normalized to the Ti $L_{2,3}$-edge intensity. SI 5 - Schematic representation of the device used to limit ice formation on the sample at low temperature.

AUTHOR INFORMATION

**Corresponding Authors**

* Vincent MAUCHAMP: vincent.mauchamp@univ-poitier.fr

* Marie-Laure DAVID: marie-laure.david@univ-poitiers.fr

* Philippe MOREAU: philippe.moreau@cnrs-imn.fr

**Present Addresses**

[§] Current affiliation : Aix Marseille Université, CNRS, PIIM, UMR 7345, Marseille, France

**Author Contributions**

The manuscript was written through contributions of all authors. All authors have given approval to the final version of the manuscript.

**Funding Sources**

This work partially pertains to the French Government program “Investissements d’Avenir” (LABEX INTERACTIFS, reference ANR-11- LABX-0017-01 and EUR INTREE, reference ANR-18-EURE-0010). This work has been partially supported by Nouvelle Aquitaine Region and by the European Structural and Investment Funds (with reference: P-2016- BAFE-94/95). The authors acknowledge financial support from the “Agence National de la Recherche” (reference ANR-24-CE09-7979, 2Dfects project).

ACKNOWLEDGMENT

Most of the transmission electron microscopy measurements were carried out with the Nant’Themis, part of IMN’s characterization platform, PLASSMAT, Nantes, France. This work partially pertains to the French Government program “Investissements d’Avenir” (LABEX INTERACTIFS, reference ANR-11- LABX-0017-01 and EUR INTREE, reference ANR-18-EURE-0010). This work has been partially supported by Nouvelle Aquitaine Region and by the European Structural and Investment Funds (with reference: P-2016- BAFE-94/95). The authors acknowledge financial support from the “Agence National de la Recherche” (reference ANR-24-CE09-7979, 2Dfects project). Dr. Clémence Queffelec from CEISAM laboratory in Nantes (France) is thanked for performing the CHNS/O analyses, as well as Jonathan Hamon (IMN laboratory, Nantes) for XPS quantification and Patrick Chartier (Pprime Institute) for the synthesis of $Ti_3AlC_2$ MAX precursors.

REFERENCES

(1) Kaiser, S. K.; Chen, Z.; Faust Akl, D.; Mitchell, S.; Pérez-Ramírez, J. Single-Atom Catalysts across the Periodic Table. *Chem. Rev.* **2020**, *120* (21), 11703–11809. https://doi.org/10.1021/acs.chemrev.0c00576

(2) Anasori, B.; Lukatskaya, M. R.; Gogotsi, Y. 2D Metal Carbides and Nitrides (MXenes) for Energy Storage. *Nat. Rev. Mater.* **2017**, *2* (2), 16098. https://doi.org/10.1038/natrevmats.2016.98

(3) Duan, X.; Wang, C.; Pan, A.; Yu, R.; Duan, X. Two-Dimensional Transition Metal Dichalcogenides as Atomically Thin Semiconductors: Opportunities and Challenges. *Chem. Soc. Rev.* **2015**, *44* (24), 8859–8876. https://doi.org/10.1039/C5CS00507H

(4) Sun, H.; Liang, J.; Zhu, J. Editorial: Interface and Structure Designs of Electrode Materials for Advanced Electrochemical Energy Storage. *Front. Energy Res.* **2023**, *11*, 1184606. https://doi.org/10.3389/fenrg.2023.1184606

(5) Mitterreiter, E.; Schuler, B.; Micevic, A.; Hernangómez-Pérez, D.; Barthelmi, K.; Cochrane, K. A.; Kiemle, J.; Sigger, F.; Klein, J.; Wong, E.; Barnard, E. S.; Watanabe, K.; Taniguchi, T.; Lorke, M.; Jahnke, F.; Finley, J. J.; Schwartzberg, A. M.; Qiu, D. Y.; Refaely-Abramson, S.; Holleitner, A. W.; Weber-Bargioni, A.; Kastl, C. The Role of Chalcogen Vacancies for Atomic Defect Emission in MoS2. *Nat. Commun.* **2021**, *12* (1), 3822. https://doi.org/10.1038/s41467-021-24102-y

(6) Zhang, X.; Han, G.; Zhu, S. Flash Nitrogen-Doped Carbon Nanotubes for Energy Storage and Conversion. *Small* **2024**, *20* (3), 2305406. https://doi.org/10.1002/smll.202305406

(7) Zhou, J.; Dahlqvist, M.; Björk, J.; Rosen, J. Atomic Scale Design of MXenes and Their Parent Materials─From Theoretical and Experimental Perspectives. *Chem. Rev.* **2023**, *123* (23), 13291–13322. https://doi.org/10.1021/acs.chemrev.3c00241

(8) Cao, Y.; Fatemi, V.; Fang, S.; Watanabe, K.; Taniguchi, T.; Kaxiras, E.; Jarillo-Herrero, P. Unconventional Superconductivity in Magic-Angle Graphene Superlattices. *Nature* **2018**, *556* (7699), 43–50. https://doi.org/10.1038/nature26160

(9) Yankowitz, M.; Chen, S.; Polshyn, H.; Zhang, Y.; Watanabe, K.; Taniguchi, T.; Graf, D.; Young, A. F.; Dean, C. R. Tuning Superconductivity in Twisted Bilayer Graphene. *Science* **2019**, *363* (6431), 1059–1064. https://doi.org/10.1126/science.aav1910

(10) Egerton, R. *Physical Principles of Electron Microscopy: An Introduction to TEM, SEM, and AEM*, 2nd ed. 2016.; SpringerLink Bücher; Springer: Cham, 2016. https://doi.org/10.1007/978-3-319-39877-8

(11) Krasheninnikov, A. V.; Banhart, F. Engineering of Nanostructured Carbon Materials with Electron or Ion Beams. *Nat. Mater.* **2007**, *6* (10), 723–733. https://doi.org/10.1038/nmat1996

(12) Bui, T. A.; Leuthner, G. T.; Madsen, J.; Monazam, M. R. A.; Chirita, A. I.; Postl, A.; Mangler, C.; Kotakoski, J.; Susi, T. Creation of Single Vacancies in hBN with Electron Irradiation. *Small* **2023**, *19* (39), 2301926. https://doi.org/10.1002/smll.202301926

(13) Kretschmer, S.; Lehnert, T.; Kaiser, U.; Krasheninnikov, A. V. Formation of Defects in Two-Dimensional $MoS_2$ in the Transmission Electron Microscope at Electron Energies below the Knock-on Threshold: The Role of Electronic Excitations. *Nano Lett.* **2020**, *20* (4), 2865–2870. https://doi.org/10.1021/acs.nanolett.0c00670

(14) Boston, R.; Schnepp, Z.; Nemoto, Y.; Sakka, Y.; Hall, S. R. In Situ TEM Observation of a Microcrucible Mechanism of Nanowire Growth. *Science* **2014**, *344* (6184), 623–626. https://doi.org/10.1126/science.1251594

(15) Gonzalez-Martinez, I. G.; Bachmatiuk, A.; Bezugly, V.; Kunstmann, J.; Gemming, T.; Liu, Z.; Cuniberti, G.; Rümmeli, M. H. Electron-Beam Induced Synthesis of Nanostructures: A Review. *Nanoscale* **2016**, *8* (22), 11340–11362. https://doi.org/10.1039/C6NR01941B

(16) Roccapriore, K. M.; Cho, S.; Lupini, A. R.; Milliron, D. J.; Kalinin, S. V. Sculpting the Plasmonic Responses of Nanoparticles by Directed Electron Beam Irradiation. *Small* **2022**, *18* (1), 2105099. https://doi.org/10.1002/smll.202105099

(17) Dyck, O.; Kim, S.; Jimenez-Izal, E.; Alexandrova, A. N.; Kalinin, S. V.; Jesse, S. Building Structures Atom by Atom via Electron Beam Manipulation. *Small* **2018**, *14* (38), 1801771. https://doi.org/10.1002/smll.201801771

(18) Song, B.; Schneider, G. F.; Xu, Q.; Pandraud, G.; Dekker, C.; Zandbergen, H. Atomic-Scale Electron-Beam Sculpting of Near-Defect-Free Graphene Nanostructures. *Nano Lett.* **2011**, *11* (6), 2247–2250. https://doi.org/10.1021/nl200369r

(19) Mendes, R. G.; Ta, H. Q.; Gemming, T.; Van Gog, H.; Van Huis, M. A.; Bachmatiuk, A.; Rümmeli, M. H. In Situ Growth of Suspended Zirconene Islets Inside Graphene Pores. *Adv. Funct. Mater.* **2025**, *35* (2), 2412889. https://doi.org/10.1002/adfm.202412889

(20) Boebinger, M. G.; Yilmaz, D. E.; Ghosh, A.; Misra, S.; Mathis, T. S.; Kalinin, S. V.; Jesse, S.; Gogotsi, Y.; Van Duin, A. C. T.; Unocic, R. R. Direct Fabrication of Atomically Defined Pores in MXenes Using Feedback-Driven STEM. *Small Methods* **2024**, 2400203. https://doi.org/10.1002/smtd.202400203

(21) David, M.-L.; Alix, K.; Pailloux, F.; Mauchamp, V.; Couillard, M.; Botton, G. A.; Pizzagalli, L. *In Situ* Controlled Modification of the Helium Density in Single Helium-Filled Nanobubbles. *J. Appl. Phys.* **2014**, *115* (12), 123508. https://doi.org/10.1063/1.4869213

(22) VahidMohammadi, A.; Rosen, J.; Gogotsi, Y. The World of Two-Dimensional Carbides and Nitrides (MXenes). *Science* **2021**, *372* (6547), eabf1581. https://doi.org/10.1126/science.abf1581

(23) Naguib, M.; Kurtoglu, M.; Presser, V.; Lu, J.; Niu, J.; Heon, M.; Hultman, L.; Gogotsi, Y.; Barsoum, M. W. Two-Dimensional Nanocrystals Produced by Exfoliation of Ti3AlC2. *Adv. Mater.* **2011**, *23* (37), 4248–4253. https://doi.org/10.1002/adma.201102306

(24) Shayesteh Zeraati, A.; Mirkhani, S. A.; Sun, P.; Naguib, M.; Braun, P. V.; Sundararaj, U. Improved Synthesis of $Ti_3C_2T_x$ MXenes Resulting in Exceptional Electrical Conductivity, High Synthesis Yield, and Enhanced Capacitance. *Nanoscale* **2021**, *13* (6), 3572–3580. https://doi.org/10.1039/D0NR06671K

(25) Maleski, K.; Shuck, C. E.; Fafarman, A. T.; Gogotsi, Y. The Broad Chromatic Range of Two-Dimensional Transition Metal Carbides. *Adv. Opt. Mater.* **2021**, *9* (4), 2001563. https://doi.org/10.1002/adom.202001563

(26) Abdolhosseinzadeh, S.; Jiang, X.; Zhang, H.; Qiu, J.; Zhang, C. (John). Perspectives on Solution Processing of Two-Dimensional MXenes. *Mater. Today* **2021**, *48*, 214–240. https://doi.org/10.1016/j.mattod.2021.02.010

(27) Hantanasirisakul, K.; Zhao, M.; Urbankowski, P.; Halim, J.; Anasori, B.; Kota, S.; Ren, C. E.; Barsoum, M. W.; Gogotsi, Y. Fabrication of $Ti_3C_2T_x$ MXene Transparent Thin Films with Tunable Optoelectronic Properties. *Adv. Electron. Mater.* **2016**, *2* (6), 1600050. https://doi.org/10.1002/aelm.201600050

(28) Zhang, C. J.; Anasori, B.; Seral-Ascaso, A.; Park, S.-H.; McEvoy, N.; Shmeliov, A.; Duesberg, G. S.; Coleman, J. N.; Gogotsi, Y.; Nicolosi, V. Transparent, Flexible, and Conductive 2D Titanium Carbide (MXene) Films with High Volumetric Capacitance. *Adv. Mater.* **2017**, *29* (36), 1702678. https://doi.org/10.1002/adma.201702678

(29) Zhang, C. (John); Ma, Y.; Zhang, X.; Abdolhosseinzadeh, S.; Sheng, H.; Lan, W.; Pakdel, A.; Heier, J.; Nüesch, F. Two-Dimensional Transition Metal Carbides and Nitrides (MXenes): Synthesis, Properties, and Electrochemical Energy Storage Applications. *ENERGY Environ. Mater.* **2020**, *3* (1), 29–55. https://doi.org/10.1002/eem2.12058

(30) Panda, S.; Deshmukh, K.; Khadheer Pasha, S. K.; Theerthagiri, J.; Manickam, S.; Choi, M. Y. MXene Based Emerging Materials for Supercapacitor Applications: Recent Advances, Challenges, and Future Perspectives. *Coord. Chem. Rev.* **2022**, *462*, 214518. https://doi.org/10.1016/j.ccr.2022.214518

(31) Huang, K.; Li, Z.; Lin, J.; Han, G.; Huang, P. Two-Dimensional Transition Metal Carbides and Nitrides (MXenes) for Biomedical Applications. *Chem. Soc. Rev.* **2018**, *47* (14), 5109–5124. https://doi.org/10.1039/C7CS00838D

(32) Soleymaniha, M.; Shahbazi, M.-A.; Rafieerad, A. R.; Maleki, A.; Amiri, A. Promoting Role of MXene Nanosheets in Biomedical Sciences: Therapeutic and Biosensing Innovations. *Adv. Healthc. Mater.* **2019**, *8* (1), 1801137. https://doi.org/10.1002/adhm.201801137

(33) Morales-Salvador, R.; Gouveia, J. D.; Morales-García, Á.; Viñes, F.; Gomes, J. R. B.; Illas, F. Carbon Capture and Usage by MXenes. *ACS Catal.* **2021**, *11* (17), 11248–11255. https://doi.org/10.1021/acscatal.1c02663

(34) Kim, S. J.; Koh, H.-J.; Ren, C. E.; Kwon, O.; Maleski, K.; Cho, S.-Y.; Anasori, B.; Kim, C.-K.; Choi, Y.-K.; Kim, J.; Gogotsi, Y.; Jung, H.-T. Metallic $Ti_3C_2T_x$ MXene Gas Sensors with Ultrahigh Signal-to-Noise Ratio. *ACS Nano* **2018**, *12* (2), 986–993. https://doi.org/10.1021/acsnano.7b07460

(35) Pei, Y.; Zhang, X.; Hui, Z.; Zhou, J.; Huang, X.; Sun, G.; Huang, W. $Ti_3C_2T_X$ MXene for Sensing Applications: Recent Progress, Design Principles, and Future Perspectives. *ACS Nano* **2021**, *15* (3), 3996–4017. https://doi.org/10.1021/acsnano.1c00248

(36) Iqbal, A.; Shahzad, F.; Hantanasirisakul, K.; Kim, M.-K.; Kwon, J.; Hong, J.; Kim, H.; Kim, D.; Gogotsi, Y.; Koo, C. M. Anomalous Absorption of Electromagnetic Waves by 2D Transition Metal Carbonitride $Ti_3CNT_x$ (MXene). *Science* **2020**, *369* (6502), 446–450. https://doi.org/10.1126/science.aba7977

(37) Shahzad, F.; Alhabeb, M.; Hatter, C. B.; Anasori, B.; Man Hong, S.; Koo, C. M.; Gogotsi, Y. Electromagnetic Interference Shielding with 2D Transition Metal Carbides (MXenes). *Science* **2016**, *353* (6304), 1137–1140. https://doi.org/10.1126/science.aag2421

(38) Sokol, M.; Natu, V.; Kota, S.; Barsoum, M. W. On the Chemical Diversity of the MAX Phases. *Trends Chem.* **2019**, *1* (2), 210–223. https://doi.org/10.1016/j.trechm.2019.02.016

(39) *2D Metal Carbides and Nitrides (MXenes): Structure, Properties and Applications*; Anasori, B., Gogotsi, Y., Eds.; Springer International Publishing: Cham, 2019. https://doi.org/10.1007/978-3-030-19026-2

(40) Luo, J.; Zhang, W.; Yuan, H.; Jin, C.; Zhang, L.; Huang, H.; Liang, C.; Xia, Y.; Zhang, J.; Gan, Y.; Tao, X. Pillared Structure Design of MXene with Ultralarge Interlayer Spacing for High-Performance Lithium-Ion Capacitors. *ACS Nano* **2017**, *11* (3), 2459–2469. https://doi.org/10.1021/acsnano.6b07668

(41) Kamysbayev, V.; Filatov, A. S.; Hu, H.; Rui, X.; Lagunas, F.; Wang, D.; Klie, R. F.; Talapin, D. V. Covalent Surface Modifications and Superconductivity of Two-Dimensional Metal Carbide MXenes. *Science* **2020**, *369* (6506), 979–983. https://doi.org/10.1126/science.aba8311

(42) Mauchamp, V.; Bugnet, M.; Bellido, E. P.; Botton, G. A.; Moreau, P.; Magne, D.; Naguib, M.; Cabioc'h, T.; Barsoum, M. W. Enhanced and Tunable Surface Plasmons in Two-Dimensional Ti 3 C 2 Stacks: Electronic Structure versus Boundary Effects. *Phys. Rev. B* **2014**, *89* (23), 235428. https://doi.org/10.1103/PhysRevB.89.235428
(43) Berdiyorov, G. R. Optical Properties of Functionalized Ti3C2T2 (T = F, O, OH) MXene: First-Principles Calculations. *AIP Adv.* **2016**, *6* (5), 055105. https://doi.org/10.1063/1.4948799
(44) Hart, J. L.; Hantanasirisakul, K.; Lang, A. C.; Anasori, B.; Pinto, D.; Pivak, Y.; Van Omme, J. T.; May, S. J.; Gogotsi, Y.; Taheri, M. L. Control of MXenes' Electronic Properties through Termination and Intercalation. *Nat. Commun.* **2019**, *10* (1), 522. https://doi.org/10.1038/s41467-018-08169-8
(45) Han, M.; Maleski, K.; Shuck, C. E.; Yang, Y.; Glazar, J. T.; Foucher, A. C.; Hantanasirisakul, K.; Sarycheva, A.; Frey, N. C.; May, S. J.; Shenoy, V. B.; Stach, E. A.; Gogotsi, Y. Tailoring Electronic and Optical Properties of MXenes through Forming Solid Solutions. *J. Am. Chem. Soc.* **2020**, *142* (45), 19110–19118. https://doi.org/10.1021/jacs.0c07395
(46) Ibragimova, R.; Lv, Z.-P.; Komsa, H.-P. First Principles Study of the Stability of MXenes under an Electron Beam. *Nanoscale Adv.* **2021**, *3* (7), 1934–1941. https://doi.org/10.1039/D0NA00886A
(47) Zhang, H.; Hu, T.; Sun, W.; Hu, M.; Cheng, R.; Wang, X. Atomic Repartition in MXenes by Electron Probes. *Chem. Mater.* **2019**, *31* (12), 4385–4391. https://doi.org/10.1021/acs.chemmater.9b00470
(48) Sang, X.; Xie, Y.; Yilmaz, D. E.; Lotfi, R.; Alhabeb, M.; Ostadhossein, A.; Anasori, B.; Sun, W.; Li, X.; Xiao, K.; Kent, P. R. C.; Van Duin, A. C. T.; Gogotsi, Y.; Unocic, R. R. In Situ Atomistic Insight into the Growth Mechanisms of Single Layer 2D Transition Metal Carbides. *Nat. Commun.* **2018**, *9* (1), 2266. https://doi.org/10.1038/s41467-018-04610-0
(49) Lagunas, F.; Zhou, C.; Wang, D.; Thakur, A.; Anasori, B.; Talapin, D. V.; Hood, Z. D.; Klie, R. F. In Situ Formation of Ripplocations in Hybrid Organic–Inorganic MXenes. *Adv. Mater.* **2025**, *37* (13), 2411669. https://doi.org/10.1002/adma.202411669
(50) Benchakar, M.; Loupias, L.; Garnero, C.; Bilyk, T.; Morais, C.; Canaff, C.; Guignard, N.; Morisset, S.; Pazniak, H.; Hurand, S.; Chartier, P.; Pacaud, J.; Mauchamp, V.; Barsoum, M. W.; Habrioux, A.; Célérier, S. One MAX Phase, Different MXenes: A Guideline to Understand the Crucial Role of Etching Conditions on Ti3C2Tx Surface Chemistry. *Appl. Surf. Sci.* **2020**, *530*, 147209. https://doi.org/10.1016/j.apsusc.2020.147209
(51) Benmoumen, A.; David, M.-L.; Gautron, E.; Hurand, S.; Tangui, S.; Loupias, L.; Habrioux, A.; Célérier, S.; Moreau, P.; Mauchamp, V. Structural and Property Engineering of 2D Titanium Carbides (MXene) Thin Films Using Ion Irradiation. *Appl. Surf. Sci.* **2024**, *652*, 159206. https://doi.org/10.1016/j.apsusc.2023.159206
(52) Malis, T.; Cheng, S. C.; Egerton, R. F. EELS Log-Ratio Technique for Specimen-Thickness Measurement in the TEM. *J. Electron Microsc. Tech.* **1988**, *8* (2), 193–200. https://doi.org/10.1002/jemt.1060080206
(53) Egerton, R. F. *Electron Energy-Loss Spectroscopy in the Electron Microscope*, 3d edition.; The language of science; Springer: New York, 2011.
(54) Bilyk, T.; Hsiao, H.-W.; Yuan, R.; Benchakar, M.; Habrioux, A.; Célérier, S.; Zuo, J.-M.; Pacaud, J.; Mauchamp, V. Plasmon Spectroscopy for the Determination of Ti $_3$ C $_2$ T $_x$ MXene Few Layer Stacks Architecture. *2D Mater.* **2022**, *9* (3), 035017. https://doi.org/10.1088/2053-1583/ac74ca
(55) Stadelmann, P. JEMS, 2004. https://www.epfl.ch/research/facilities/cime/research/research-jems/

(56) Pazniak, H.; Benchakar, M.; Bilyk, T.; Liedl, A.; Busby, Y.; Noël, C.; Chartier, P.; Hurand, S.; Marteau, M.; Houssiau, L.; Larciprete, R.; Lacovig, P.; Lizzit, D.; Tosi, E.; Lizzit, S.; Pacaud, J.; Célérier, S.; Mauchamp, V.; David, M.-L. Ion Implantation as an Approach for Structural Modifications and Functionalization of $Ti_3C_2T_x$ MXenes. *ACS Nano* **2021**, *15* (3), 4245–4255. https://doi.org/10.1021/acsnano.0c06735

(57) Dessoliers, Z.; Chemin, A.; Valurouthu, G.; Lord, R.; Bilyk, T.; Gogotsi, Y.; Mauchamp, V.; Petit, T. Combining X-Ray Photoelectron and Absorption Spectroscopies for Determining Surface Chemistry and Composition of $Ti_3C_2T_x$ MXene. *Adv. Mater. Interfaces* **2025**, *12* (13), 2500391. https://doi.org/10.1002/admi.202500391

(58) Mauchamp, V.; Bugnet, M.; Bilyk, T.; Huot, C.; Célérier, S.; Massin, L.; Gelin, P.; Aouine, M.; Pacaud, J. Selective Gas Response of MXene Surface Functional Groups Revealed by Gas-Phase Transmission Electron Microscopy. *ACS Nano* **2026**, *20* (12), 9828–9840. https://doi.org/10.1021/acsnano.5c19308

(59) Lu, J.; Persson, I.; Lind, H.; Palisaitis, J.; Li, M.; Li, Y.; Chen, K.; Zhou, J.; Du, S.; Chai, Z.; Huang, Z.; Hultman, L.; Eklund, P.; Rosen, J.; Huang, Q.; Persson, P. O. Å. $Ti_{n+1}C_n$ MXenes with Fully Saturated and Thermally Stable Cl Terminations. *Nanoscale Adv.* **2019**, *1* (9), 3680–3685. https://doi.org/10.1039/C9NA00324J

(60) Li, Y.; Shao, H.; Lin, Z.; Lu, J.; Liu, L.; Duployer, B.; Persson, P. O. Å.; Eklund, P.; Hultman, L.; Li, M.; Chen, K.; Zha, X.-H.; Du, S.; Rozier, P.; Chai, Z.; Raymundo-Piñero, E.; Taberna, P.-L.; Simon, P.; Huang, Q. A General Lewis Acidic Etching Route for Preparing MXenes with Enhanced Electrochemical Performance in Non-Aqueous Electrolyte. *Nat. Mater.* **2020**, *19* (8), 894–899. https://doi.org/10.1038/s41563-020-0657-0

(61) Li, M.; Lu, J.; Luo, K.; Li, Y.; Chang, K.; Chen, K.; Zhou, J.; Rosen, J.; Hultman, L.; Eklund, P.; Persson, P. O. Å.; Du, S.; Chai, Z.; Huang, Z.; Huang, Q. Element Replacement Approach by Reaction with Lewis Acidic Molten Salts to Synthesize Nanolaminated MAX Phases and MXenes. *J. Am. Chem. Soc.* **2019**, *141* (11), 4730–4737. https://doi.org/10.1021/jacs.9b00574

(62) Persson, I.; Halim, J.; Lind, H.; Hansen, T. W.; Wagner, J. B.; Näslund, L.; Darakchieva, V.; Palisaitis, J.; Rosen, J.; Persson, P. O. Å. 2D Transition Metal Carbides (MXenes) for Carbon Capture. *Adv. Mater.* **2019**, *31* (2), 1805472. https://doi.org/10.1002/adma.201805472

(63) Koh, S. W.; Rekhi, L.; Arramel; Birowosuto, M. D.; Trinh, Q. T.; Ge, J.; Yu, W.; Wee, A. T. S.; Choksi, T. S.; Li, H. Tuning the Work Function of MXene via Surface Functionalization. *ACS Appl. Mater. Interfaces* **2024**, *16* (49), 66826–66836. https://doi.org/10.1021/acsami.3c11857

(64) Khazaei, M.; Arai, M.; Sasaki, T.; Ranjbar, A.; Liang, Y.; Yunoki, S. OH-Terminated Two-Dimensional Transition Metal Carbides and Nitrides as Ultralow Work Function Materials. *Phys. Rev. B* **2015**, *92* (7), 075411. https://doi.org/10.1103/PhysRevB.92.075411

(65) Lu, C.; Yang, L.; Yan, B.; Sun, L.; Zhang, P.; Zhang, W.; Sun, Z. Nitrogen-Doped $Ti_3C_2$ MXene: Mechanism Investigation and Electrochemical Analysis. *Adv. Funct. Mater.* **2020**, *30* (47), 2000852. https://doi.org/10.1002/adfm.202000852

(66) Bilyk, T.; Benchakar, M.; Bugnet, M.; Loupias, L.; Chartier, P.; Pazniak, H.; David, M.-L.; Habrioux, A.; Celerier, S.; Pacaud, J.; Mauchamp, V. Electronic Structure Sensitivity to Surface Disorder and Nanometer-Scale Impurity of 2D Titanium Carbide MXene Sheets as Revealed by Electron Energy-Loss Spectroscopy. *J. Phys. Chem. C* **2020**, *124* (49), 27071–27081. https://doi.org/10.1021/acs.jpcc.0c06798

(67) Huang, X.; Feng, Y.; Dou, Y.; Tang, H.; Ding, D.; Tian, P.; Xia, M.; Qian, G.; Wang, Y.; Zhang, X. Effect of Electron Irradiation on Ti 3 AlC 2. *Scr. Mater.* **2016**, *113*, 114–117. https://doi.org/10.1016/j.scriptamat.2015.10.031
(68) Garcia, V.; Glachant, A.; Pantel, R.; Straboni, A. Electron-Beam-Induced Nitrogen Migration through a Nitrided Silicon Dioxide Thin Film. *Appl. Surf. Sci.* **1994**, *74* (2), 165–170. https://doi.org/10.1016/0169-4332(94)90298-4
(69) Lam, N. Q.; Wiedersich, H. Bombardment-Induced Segregation and Redistribution. *Nucl. Instrum. Methods Phys. Res. Sect. B Beam Interact. Mater. At.* **1986**, *18* (1–6), 471–485. https://doi.org/10.1016/S0168-583X(86)80073-8
(70) Lam, N. Q.; Okamoto, P. R.; Johnson, R. A. Solute Segregation and Precipitation under Heavy-Ion Bombardment. *J. Nucl. Mater.* **1978**, *78* (2), 408–418. https://doi.org/10.1016/0022-3115(78)90462-2
(71) Rehn, L. E.; Okamoto, P. R. Production of Freely-Migrating Defects during Irradiation. *Mater. Sci. Forum* **1987**, *15–18*, 985–1002. https://doi.org/10.4028/www.scientific.net/MSF.15-18.985
(72) Moncoffre, N.; Jagielski, J. Redistribution of Nitrogen Implanted into Iron: Role of Carbon and Radiation Defects. *Surf. Coat. Technol.* **1994**, *65* (1–3), 30–36. https://doi.org/10.1016/S0257-8972(94)80005-7
(73) Wyatt, B. C.; Rosenkranz, A.; Anasori, B. 2D MXenes: Tunable Mechanical and Tribological Properties. *Adv. Mater.* **2021**, *33* (17), 2007973. https://doi.org/10.1002/adma.202007973